# Robust Faster-than-Nyquist PDM-mQAM Systems with Tomlinson-Harashima Precoding

Deyuan Chang, Oluyemi Omomukuyo, Xiang Lin, Shu Zhang,
Octavia A. Dobre, and Ramachandran Venkatesan

*Abstract*—A training-based channel estimation algorithm is proposed for the faster-than-Nyquist PDM-mQAM (m = 4, 16, 64) systems with Tomlinson-Harashima precoding (THP). This is robust to the convergence failure phenomenon suffered by the existing algorithm, yet remaining format-transparent. Simulation results show that the proposed algorithm requires a reduced optical signal-to-noise ratio (OSNR) to achieve a certain bit error rate (BER) in the presence of first-order polarization mode dispersion and phase noise introduced by the laser linewidth.

## I. INTRODUCTION

FOR the future software-defined elastic optical networking, it is highly desirable that the spectral efficiency is adjusted according to the optical channel conditions by modifying the symbol rate, the modulation format and so on [1]. An efficient way of adjusting the spectral efficiency is faster-than-Nyquist (FTN) signaling, which has attracted intensive investigations in recent years [2]-[5] and provides more flexibility for the optical networking. FTN tightly packs multi-channel signals, each with a channel spacing narrower than the symbol rate, by introducing controlled inter-carrier or inter-symbol interference (ICI/ISI).

FTN is beneficial in some application scenarios with the specific channel grid restriction. It has been experimentally demonstrated [6] that for 25 GHz spaced systems with an identical bit rate per channel of 112 Gbit/s, the polarization division multiplexed (PDM) quadrature phase shift keying (QPSK) modulated FTN scheme provides a 3.4 dB increased sensitivity than the conventional PDM 16-ary quadrature amplitude modulation (16-QAM) in the back-to-back configuration.

In order to handle the ICI/ISI impairment in the PDM-QPSK modulated FTN systems, several techniques were proposed, such as the maximum likelihood sequence estimation (MLSE) and the decision feedback equalization (DFE) [2]-[6]. However, an important drawback of these techniques is that they have to be implemented after the butterfly-type adaptive equalizers for polarization de-multiplexing, and as a result, the adaptive equalizers need to be tolerant to the introduced ICI/ISI to achieve the convergence. As the order of the modulation increases to 16-QAM, the introduced ICI/ISI leads to an indistinguishable constellation diagram because of the reduction in the Euclidean distance, which prevents the adaptive equalizers from obtaining the convergence. Hence, the above techniques proposed for the FTN-PDM-QPSK system are not suitable for the FTN-PDM-16QAM systems. In [5], the authors set up an FTN-PDM-16QAM system by using the MLSE-based technique, and the convergence failures were experimentally observed.

For the first time, we demonstrated the feasibility of the FTN-DP-16QAM system using Tomlinson-Harashima precoding (THP) in [7]. The feedforward equalization (FFE) part of THP was located before the polarization de-multiplexing module, such that the impact of the ISI caused by the narrowband filtering on the adaptive algorithm was alleviated. THP expands the original constellation points to additional points outside the modulo boundary [8]. As a result, a training symbol (TS) no longer has a deterministic location, which is harmful to the convergence of the training-based adaptive algorithm. Fortunately, when compared with other constellation points, the four points on the innermost circle of the 16-QAM constellation have a significantly reduced probability to be expanded; this makes them suitable TS candidates to be employed in the traditional training symbol least mean squares (TS-LMS) algorithm. Nonetheless, as shown in Section III, the TS-LMS still has a relatively high probability of convergence failure in the presence of first-order polarization mode dispersion (PMD), or so called differential group delay (DGD) and phase noise introduced by the laser linewidth. Furthermore, with the increase of the DGD value, the convergence failure rate increases.

In this letter, we propose a modified training-based channel estimation adaptive algorithm by taking the expanded constellation points into account. According to the simulation results, it enables the THP coded FTN-PDM-mQAM systems being robust in the presence of the DGD effect, namely, it completely eliminates the convergence failure phenomenon. Additionally, when compared with the existing algorithm, the required OSNR values to reach the BER of $2.42 \times 10^{-2}$ are reduced by 0.6 dB, 0.2 dB and 0.6 dB for 4-, 16- and 64-QAM in FTN systems, respectively, in the presence of first-order PMD and phase noise introduced by the laser linewidth.

This work was supported in part by the Atlantic Canada Opportunities Agency (ACOA), in part by the Research and Development Corporation (RDC).

D. Chang, O. Omomukuyo, X. Lin, S. Zhang, O. Dobre, and R. Venkatesan are with the Faculty of Engineering and Applied Science, Memorial University, St. John's, NL, A1B 3X5, Canada (e-mail: dchang@mun.ca).

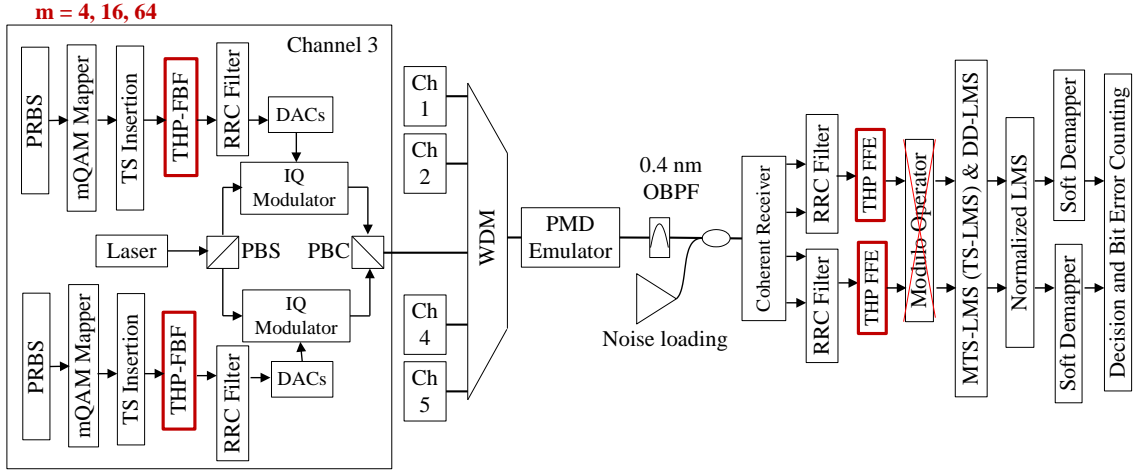

Fig. 1. The schematic diagram of the FTN-PDM-mQAM system (RRC: root-raised cosine, PBS: polarization beam splitter, PBC: polarization beam combiner, DAC: digital-to-analog converter, OBPF: optical band pass filter, ADC: analog-to-digital converter).

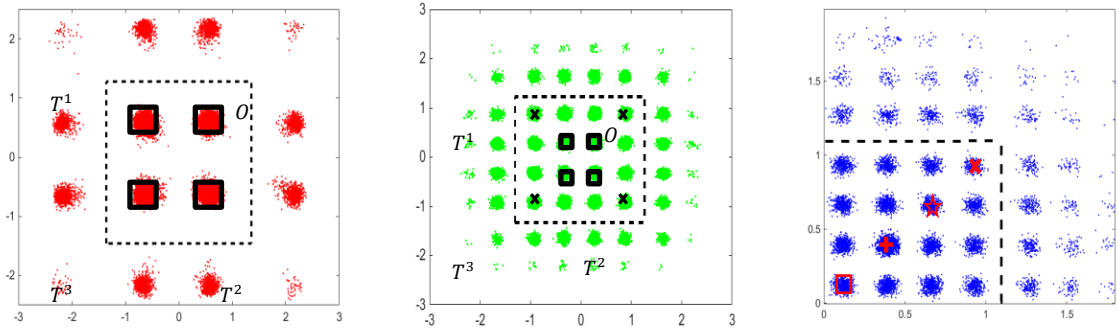

Fig. 2. The constellation diagrams for the mQAM with the THP (red color: 4-QAM; green color: 16-QAM; blue color: 64-QAM (showing only one quadrant)) with the TS types illustrated (square symbols for Type I; x-marked symbols for Type II; plus symbols for Type III; star symbols for Type IV).

Meanwhile it remains format-transparent which is especially beneficial to the elastic optical networking in terms of implementation complexity.

## II. SYSTEM SETUP AND OPERATION PRINCIPLE

### A. FTN-PDM-mQAM System Setup

Fig. 1 shows the schematic diagram of the FTN-PDM-mQAM system. Five channels are simulated with a channel spacing of 30 GHz and the BER is assessed on the central channel. The symbol rate of each channel is 32 Gbaud. Two independent pseudo random bit sequences are generated for the two polarization multiplexed branches. After the mQAM mapper on each branch, 1000 TS's are inserted at the beginning of the transmitted symbols in order to achieve the pre-convergence by using the training-based adaptive channel estimation algorithm at the receiver side. Then, 24 TS's are inserted every 1000 transmitted symbols to capture the fast dynamic channel behaviors, such as the phase cycle slip and the transient polarization state change. Then, the symbols are processed by the feedback filter of THP (THP-FBF) with the same coefficients as in [7] and the modulo size factor equal to 0.3, up-sampled by a factor of 2 and digitally shaped using the RRC filter with 73 taps, the roll-off factor of 0.1 and the 3-dB bandwidth $\Omega$ = 28.5 GHz. The polarization multiplexed optical signal is generated by the combination of a laser, a PBS, a PBC, two ideal IQ modulators and four ideal DACs. To simplify the investigation, both chromatic dispersion and nonlinearity effects are neglected, and a first-order PMD emulator is employed to investigate the performance of the adaptive channel estimation algorithms. The input state of polarization into the PMD emulator is set to 45º offset with respect to the transmitted signal to ensure the convergence performance results being obtained for the worst-case alignment case. At the receiver, the central channel is selected by using a 0.4 nm OBPF and then coherently detected by the combination of a local laser with zero frequency offset to the transmitter laser of the central channel, and a 90º hybrid. The coherently detected signal is sampled at 2 samples/symbol by ADCs and processed by the matched RRC filter. Then, after being down-sampled to 1 sample per symbol, the signal is processed by the THP-FFE, after which the traditional modulo operation of THP is removed so as to provide accurate soft information to the potential subsequent soft decision decoder [7]. The coefficients of the THP-FFE are the same as that in [7]. A symbol spaced butterfly-type adaptive filter with 11 taps is employed to compensate for the polarization effects. Here we adopt the basic structure of the phase dependent filter update algorithm proposed in [9], where the normalized LMS is used to estimate the carrier phase. In the training symbols section, either the modified TS-LMS (MTS-LMS) or the traditional TS-LMS is xused; in the data payload section, the update algorithm switches from the TS mode to the decision directed (DD) mode. This scheme can be easily applied to high-order QAM modulation formats. The bit errors are counted after the soft demapper for the THP [7]. As a comparison, Nyquist PDM-mQAM system realizations are also considered by setting $\Omega$ to 32 GHz and bypassing the THP parts, where there is overlapped bandwidth between the five 30 GHz spaced channels and, thus, the ICI exists.

*B. MTS-LMS Algorithm*

The THP operation expands the original constellation points to the additional points outside the modulo boundary, as shown in Fig. 2. When each TS is received and processed by the THP-FFE, it may be located at one of 4 possible locations: the original point $O$ with a high probability and three expanded points $T^1$, $T^2$, and $T^3$ with relatively low probabilities. In the $x$- and $y$-polarizations multiplexed systems, the expanded points will occasionally cause the convergence failures of the traditional TS-LMS algorithm, as shown in Section III.

Instead of directly using the original point $O$ in the TS-LMS algorithm, in our modified TS-LMS (MTS-LMS) algorithm we alternatively choose the TS's location $d_p(n)$, $p = x, y$, between the four possible candidate positions by the following rule:

$$d_p(n) = \arg \min_{S_p(n)} |S_p(n)(|\varphi_p(n)|/\varphi_p(n)) - E'_p(n)| \quad (1)$$

where $S_p(n) = O_p(n), T_p^1(n), T_p^2(n), T_p^3(n)$, $E'_p(n)$ represents the $n$-th symbol in the polarization branch $p$ at the output of the butterfly equalizer, and $\varphi_p(n)$ is a complex number for the $n$-th symbol in the polarization branch $p$ to compensate for the carrier phase. It's obvious that the MTS-LMS remains transparent to all the modulation formats. It can be easily figured out from Fig. 2 that for the FTN-PDM-4QAM system, there is only one TS type. The FTN-PDM-16QAM system has three TS types, which have been investigated in [10] using the traditional TS-LMS algorithm. On the other hand, there are many TS types for the FTN-PDM-64QAM system because of the large constellation size. To simplify the investigation, here we consider the equivalent constellation points along the diagonal as the TS types. As shown in Fig. 2, there are 1, 2 and 4 TS types respectively for the 4-, 16- and 64-QAM formats in the FTN systems with THP. For each TS type, the TS sequence is generated by randomly selecting the points among the corresponding four points.

## III. RESULTS

As the first step, we investigate the performance of the TS types listed above for the FTN systems using the MTS-LMS algorithm. For comparison, the performance of the 30 GHz spaced Nyquist PDM-mQAM systems without THP is also illustrated. As shown in Fig. 3, in the presence of the DGD and the laser linewidth, both the FTN-PDM-4QAM and the 5-channel Nyquist PDM-4QAM converge, although the latter has a relative flatter performance curve due to the ICI effect. The precoding loss for the 4-QAM format, namely the gap between the FTN-PDM-4QAM and the single channel Nyquist PDM-4QAM, is nearly 2.5 dB at a BER of 2.42×10$^{-2}$, which matches the theoretical result in [8]. For the FTN-PDM-16QAM and FTN-PDM-64QAM systems, it is worth notifying that only TS Types I and III, respectively, can achieve convergence in the presence of the DGD and the laser linewidth. Other TS Types lead to a BER of 0.1 even when the OSNR value is sufficiently high (for the clarity of the presentation, in Fig. 3 we do not depict all the bad results, only illustrating Type II for 16-QAM and Type I for 64-QAM, respectively). Thus, in the following simulations, Types I and III are employed respectively for the 16- and 64-QAM modulated FTN systems. Accordingly, the precoding loss for both the 16- and 64-QAM is about 1.8 dB. Additionally, due to the severe impact of the ICI effect in the high-order modulated systems, the relevant 5-channel Nyquist PDM-mQAM (m = 16, 64) systems cannot converge and have much flatter performance curves, which prohibits the subsequent probable processing, such as MLSE or DFE, to realize FTN systems. Particularly, it is easy to implement THP due to its linear filter structure and the inherent format-transparent feature [8].

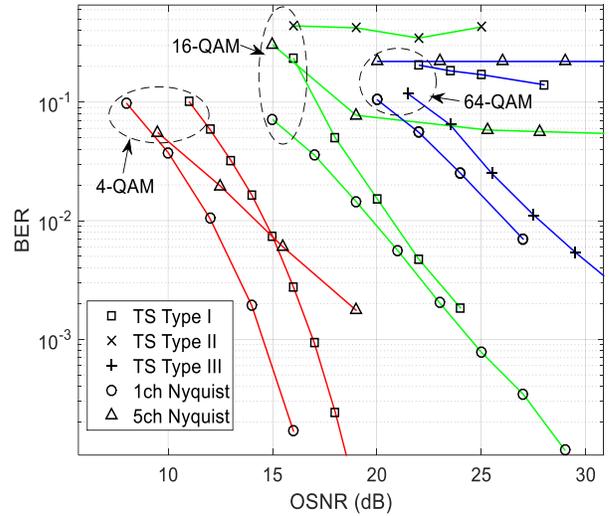

Fig. 3. Performance curves of the 5-channel FTN-PDM-mQAM using different TS types, the single channel (1ch) and the 5-channel (5ch) Nyquist PDM-mQAM systems at the laser linewidth of 10 kHz, and the DGD is 10 ps.

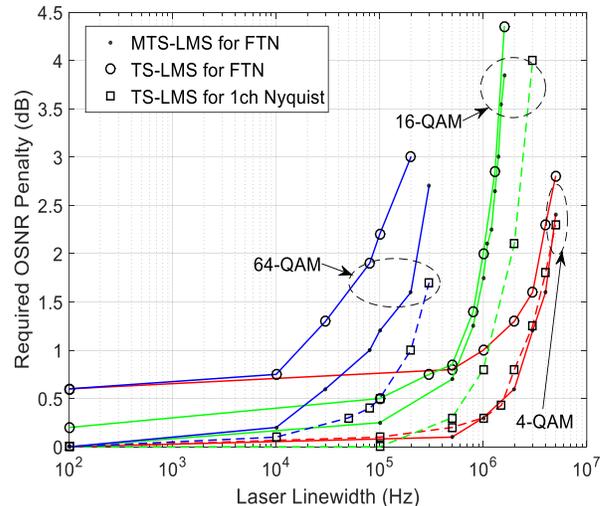

Fig. 4. Required OSNR penalty for BER = 2.42×10$^{-2}$ of the single channel Nyquist PDM-mQAM and the FTN-PDM-mQAM systems using the TS-LMS and the MTS-LMS algorithms as a function of the laser linewidth.

Hence, it can be inferred that for the elastic FTN systems, the THP-based scheme represents a very promising solution when considering both performance and implementation complexity.

We investigate the linewidth tolerance for the FTN systems in the back-to-back case. When the laser linewidth is equal to zero and the MTS-LMS algorithm is used, the required OSNR values for the FTN systems are calculated as the reference values to draw the penalty curves of the FTN systems. As shown in Fig. 4, when the TS-LMS is used, for the FTN-PDM-4QAM and FTN-PDM-64QAM systems, the penalty curves are degraded by 0.6 dB compared with those obtained by using the MTS-LMS algorithm, whereas for the FTN- PDM-16QAM, the degradation is about 0.2 dB. This can be partly explained by the observation from Fig. 2 that the expansion probability of the TS employed for the FTN-PDM-16QAM is a bit lower than that of the TS for both the FTN-PDM-4QAM and the FTN-PDM-64QAM systems, and the MTS-LMS algorithm can diminish the impact of the expanded points significantly. The penalty curves of the single channel Nyquist PDM-mQAM systems are also shown in Fig. 4. Given 1 dB required OSNR penalty, the linewidth tolerance value of 2.8 MHz for the FTN-PDM-4QAM is almost the same as that of their Nyquist counterpart systems. On the other hand, the linewidth tolerance values of 1.1 MHz and 200 kHz for the 16-QAM and 64-QAM modulated Nyquist systems are respectively decreased to 700 kHz and 80 kHz for the FTN systems enabled by THP.

Fig. 5 shows the DGD tolerance for the 4- and 16-QAM modulated FTN dual polarization systems when the laser linewidth value is fixed (note that the required OSNR value for the TS-LMS algorithm is calculated by excluding the convergence failure cases). The convergence failure rate curve is also obtained by running simulations for 10,000 trials for each given DGD value. Despite the occasional convergence failures of the TS-LMS, both the TS-LMS and the MTS-LMS can compensate for the DGD value up to 325 ps. However, for the FTN-PDM-64QAM system, the TS-LMS does not work even when the DGD value is as low as 3 ps. It is worth emphasizing that the proposed MTS-LMS converges over all simulation cases.

IV. CONCLUSION

In this letter, we propose and numerically investigate a modified training based channel estimation algorithm for the FTN-PDM-mQAM systems enabled by THP for m = 4, 16 and 64. This improves the system performance by 0.6 dB, 0.2 dB, and 0.6 dB for 4-, 16-, and 64-QAM modulated FTN systems, respectively. Additionally, it vitally eliminates the convergence failure phenomenon caused by the expanded points by choosing one from the four TS candidate locations, which makes the THP-based FTN scheme promising for the future elastic network.

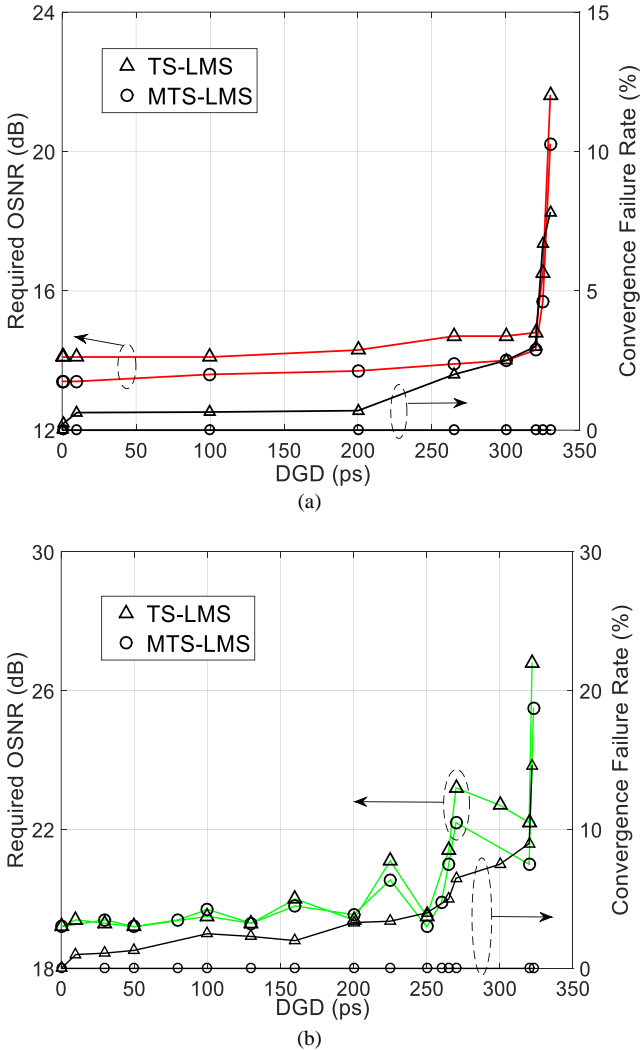

Fig. 5. Required OSNR for BER = $2.42 \times 10^{-2}$ and the convergence failure rates as a function of DGD. (a) FTN-PDM-4QAM with the linewidth of 800 kHz; (b) FTN-PDM-16QAM with the linewidth of 300 kHz.